\documentclass[reprint,english,aps,prb,raggedbottom,showpacs,floatfix]{revtex4-1}
\usepackage[T1]{fontenc}
\usepackage[utf8]{inputenc}
\usepackage{babel}
\usepackage{verbatim}
\usepackage{amsmath}
\usepackage{amssymb}
\usepackage{graphicx}
\usepackage{esint}
\usepackage{hyperref}
\usepackage{breakurl}
\usepackage[normalem]{ulem}

\usepackage[usenames, dvipsnames]{color}
\usepackage{soul}

\begin{document}

\title{Low quasiparticle coherence temperature in the one-band Hubbard model: A slave-boson approach}

\author{Alejandro Mezio}
\email{a.mezio@uq.edu.au}
\affiliation{School of Mathematics and Physics, University of Queensland, Brisbane QLD 4072, Australia}

\author{Ross H. McKenzie}
\email{r.mckenzie@uq.edu.au}
\homepage{condensedconcepts.blogspot.com.au}
\affiliation{School of Mathematics and Physics, University of Queensland, Brisbane QLD 4072, Australia}

\date{\today}
\begin{abstract}
We use the Kotliar-Ruckenstein slave-boson formalism to study the temperature dependence of paramagnetic phases of the one-band Hubbard model for a variety of band structures.
We calculate the Fermi liquid quasiparticle spectral weight $Z$ and identify the temperature at which it decreases significantly to a crossover to a bad metal region.
Near the Mott metal-insulator transition, this coherence temperature $T_\textrm{coh}$ is much lower than the Fermi temperature of the uncorrelated Fermi gas, as is observed in a broad range of strongly correlated electron materials.
After a proper rescaling of temperature and interaction, we find a universal behavior that is independent of the band structure of the system.
We obtain the temperature-interaction phase diagram as function of doping, and we compare the temperature dependence of the double occupancy, entropy, and charge compressibility with previous results obtained with Dynamical Mean-Field Theory.
We analyse the stability of the method by calculating the charge compressibility.
\end{abstract}

%\pacs{PACS, PACS, PACS, PACS}

\maketitle

\section{Introduction}

Understanding strongly correlated electron materials is a significant theoretical challenge because they exhibit a variety of phases with exotic properties. A wide range of materials exhibits this behaviour, ranging from the transition metal oxides,\cite{Imada1998} 
high-$T_c$ cuprates superconductors,\cite{Armitage2010,Keimer2015} 
and heavy fermions compounds,\cite{Degiorgi1999,Stewart2001,Lohneysen2007,Gegenwart2008}
to the organic charge transfer salts\cite{Powell2011}
and iron based superconductors.\cite{Qazilbash2009,Mazin2010,Georges2013a}
In the metallic phase some have low-temperature properties consistent with the Landau Fermi liquid (FL) picture of conventional metals
below a low energy scale, defined as coherence temperature $T_\textrm{coh}$.
This low-temperature scale signals the breakdown of the Fermi liquid picture and is orders of magnitude smaller than $T_\textrm{F}^0$, the Fermi temperature associated with the band structure for uncorrelated electrons.
%($T_\textrm{F} \simeq \frac{E_\textrm{F}}{k_\textrm{B}}$).
The family of organic salts $\kappa$-(BEDT-TTF)$_2$X has $T_\textrm{coh} \simeq 30-50\ $K, and a Fermi temperature of $T_\textrm{F}^0 \simeq 600\ $K.\cite{Powell2011}
%Also, 
Sr$_2$RuO$_4$ has $T_\textrm{coh} \simeq 30-100\ $K\cite{Tyler1998,Noce1999,Georges2013a} and LiV$_2$O$_4$ has $T_\textrm{coh} \simeq 20-30\ $K,\cite{Urano2000} both with $T_\textrm{F}^0 \sim 10^4\ $K.
%This low-temperature, signals the breakdown of the Fermi liquid picture. The family of organic salts $\kappa$-(BEDT-TTF)$_2$X has $T_\textrm{coh} \simeq 50\ $K,\cite{Powell2011} Sr$_2$RuO$_4$ has $T_\textrm{coh} \simeq 20\ $K\cite{Tyler1998} and LiV$_2$O$_4$ has $T_\textrm{coh} \simeq 20-30\ $K.\cite{Urano2000}
%These temperature scales are orders of magnitude smaller than the Fermi temperature associated with the band structure for uncorrelated electrons.
Above this coherence temperature, a transfer of spectral weight to higher energies occurs and quasi-particles do not exist.
These exotic states and the small coherence temperature are associated with the proximity to a Mott metal-insulator transition\cite{Imada1998} (MIT) or a quantum critical point.\cite{Gegenwart2008,Smith2008,Dasari2016}

Below $T_\textrm{coh}$ the transport properties can be characterised by diffusive transport of coherent quasiparticle states, where the mean-free path is much larger than the lattice constant.
In this regime, the resistivity often behaves as $\rho \propto T^2$ and is much less than the Mott-Ioffe-Regel (MIR) limit, $\frac{h a}{e^2} \sim 250\ \mu\, \Omega\, cm$, where $a$ is the lattice constant.
However, above $T_\textrm{coh}$ quasi-particles do not exist and the resistivity exceeds the MIR limit (the associated mean-free path would be smaller than the lattice spacing).
Hence, in these bad metallic states, Boltzmann transport theory breaks down and a theoretical description is particularly challenging.\cite{Emery1995}
Other signatures of the crossover from a FL to a bad metal above $T_\textrm{coh}$ are: an incoherent electron spectral function, collapse of the Drude peak in the optical conductivity and shift of the associated spectral weight to higher frequencies, the entropy and specific heat becomes of order $k_B$ per particle, the thermopower becomes of about $k_B/e$ which is orders of magnitude larger than for elemental metals, and sometimes a non-monotonic temperature dependence of the Hall constant and thermoelectric power.\cite{Rozenberg1995,Merino2000b,Merino2000,Gunnarsson2003,Hussey2004a,Merino2008,Deng2013a,Xu2013,Vucicevic2013,Vucicevic2015,Dasari2016}

The Hubbard model is the simplest Hamiltonian that captures the essential physics of the Mott MIT.
Significant theoretical progress has been made in the past two decades using Dynamical Mean-Field Theory (DMFT).\cite{Georges1996}
This method provided a detailed picture of the evolution of the electronic structure with temperature and interaction strength.
Despite being exact only in the limit of either infinite lattice connectivity or spatial dimensionality, it has been found to give a good description of three-dimensional transition metal oxides\citep{Kotliar2004} and has been argued to be relevant for the properties of two dimensional organic charge transfer salts.\cite{Powell2011,Merino2000,Merino2008}
\begin{figure}[hbt]
\includegraphics*[width=0.46\textwidth,angle=0]{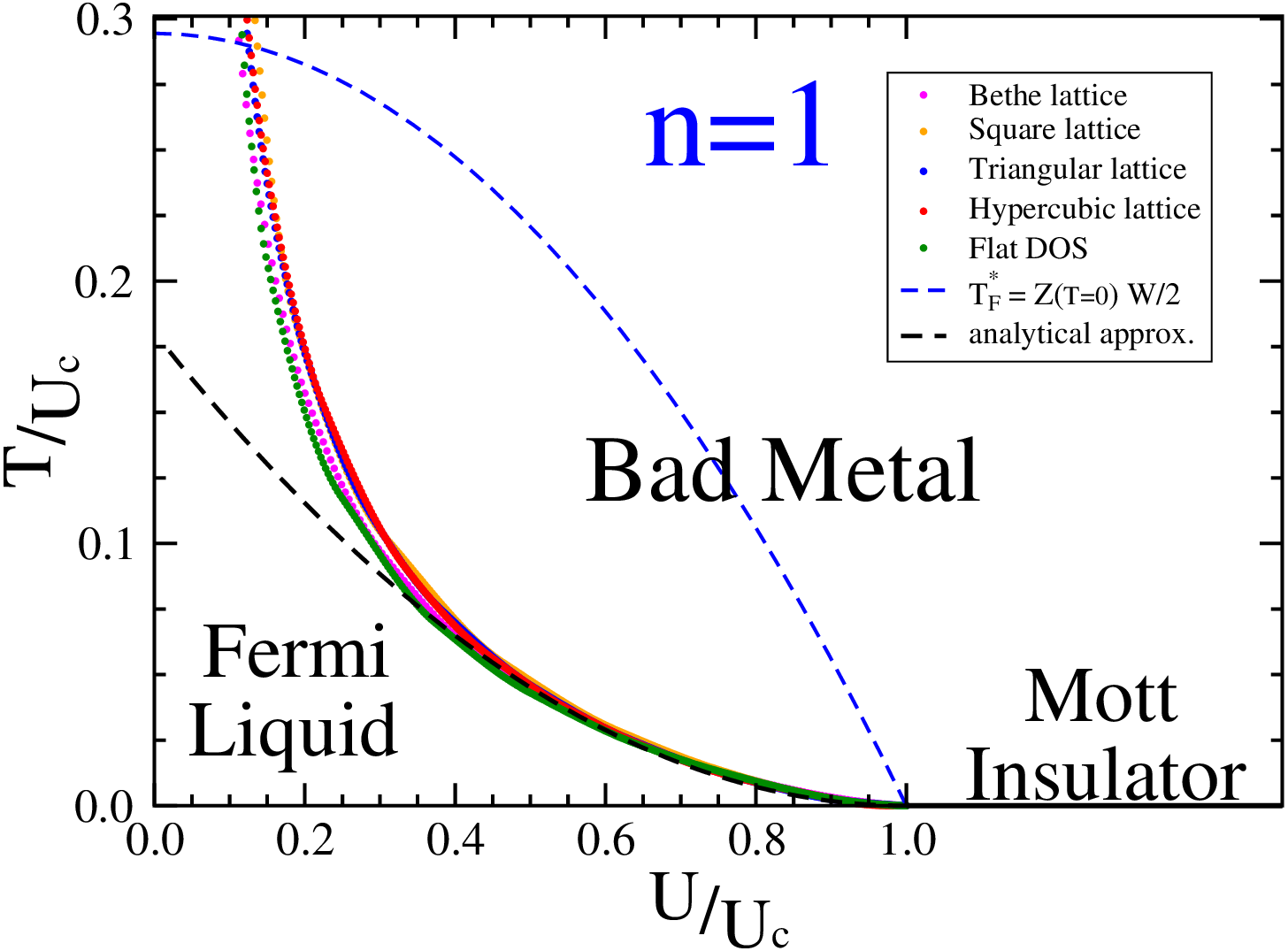}
\protect\caption{Temperature vs. interaction phase diagram for the Hubbard model at half-filling for different band structures.
We observe universal behaviour, in that the results are independent of the details of the band structure when $T$ and $U$ are scaled by $U_c$ (defined in equation \eqref{Ucritic}).
$T_{\textrm{coh}}$ (color points) is the coherence temperature, where the quasi-particle weight $Z=q$ goes to zero, calculated with the slave-boson method (cf. Fig. \ref{n-100_Z}). At $T=0$ and for $U>U_c$ the system is a Mott insulator. For $U<U_c$ and $T<T_{\textrm{coh}}$ the system is in a Fermi liquid phase. For $T>T_\textrm{coh}$ the system is in the bad metal regime.
Close to the metal-insulator transition the coherence temperature is orders of magnitude smaller than the energy scale $U_c$, which is of the order of the energy of the uncorrelated system. There is a very good agreement with the analytic approximation (dashed black line) found in Ref. [\onlinecite{Fresard1997}] (cf. equation \eqref{FKequation}).
Note also that $T_{\textrm{coh}} \ll T^*_\textrm{F} \equiv Z(T\!=\!0)\, \varepsilon_\textrm{F}^0 = Z(T\!=\!0)\, \frac{W}{2}$ (dashed blue line), where $\varepsilon_\textrm{F}^0$ is the uncorrelated Fermi energy and $W$ the bandwidth. This shows that the large reduction in $T_{\textrm{coh}}$ is not just due to the renormalization of the bandwidth.
Our results are qualitatively consistent with DMFT calculations.\cite{Rozenberg1994c,Vucicevic2013}
\label{n-100_Tcoh}}
\end{figure}

In addition to the cases of infinite dimension $d=\infty$ (exact within DMFT\cite{Georges1996}), exact solutions for the Hubbard model are only known for $d=1$,\cite{Lieb1968} the Nagaoka limit,\cite{Nagaoka1966} and the trivial case $U=0$. Also, diverse numerically controlled and numerically exact results were achieved in recent years (for a summary and careful comparison for the square lattice see Ref. [\onlinecite{Leblanc2015}] and references therein).
In terms of analytical methods, the slave-boson method introduced by Kotliar and Ruckenstein\cite{Kotliar1986} allows a non-perturbative treatment.
Since this method is an exact mapping of the electron operator, approximations are still necessary. Already at the mean-field level, the paramagnetic solution reproduces the Gutzwiller approximation.\cite{Gebhard1990}
In addition, the approach is exact in the large degeneracy limit, and it obeys a variational principle in the limit of large spatial dimensions, where the Gutzwiller approximation and the Gutzwiller wave function are identical.\cite{Metzner1989a}
These formal properties signify that the approach captures characteristic features of strongly correlated electron systems such as the suppression of the quasi-particle weight and the Brinkman-Rice picture at the MIT.\cite{Brinkman1970} For a general review of the method see Ref. [\onlinecite{Fresard2012a},\onlinecite{Gebhard2000}].

Several studies have used this method (and variants) in the study of the one band Hubbard model at finite temperature,\cite{Hasegawa1989,Hasegawa1990a,Lilly1990,Fresard,Fresard1997,Kaga2003,Camjayi2007,Lanata2015a,Dao2017}
most focused on magnetic solutions. For paramagnetic solutions and generic filling it has been found that the saddle-point equation possesses three solutions,\cite{Fresard} the physical one being that which has minimal free energy.
At half-filling, a first-order transition was found where the coherent solution ceases to exist, and an approximate analytic expression for $T_\textrm{coh}$ was calculated\cite{Fresard1997} (see equation \eqref{FKequation} below).
In a very recent work, the first-order transition was studied for the square lattice, and the stability of the phase was analysed by the spin and charge dynamical susceptibilities calculated using Gaussian fluctuations from the saddle-point solution.\cite{Dao2017}
In a somewhat similar vein to our work, previous studies of the Anderson impurity model using the Barnes-Coleman slave-boson representation identified a phase transition in the slave-boson field with the Kondo temperature.\cite{Newns1987}
Also, a large-$\mathcal{N}$ mean-field study of the Kondo-lattice model relates the Kondo temperature with the vanishing of a Hubbard-Stratonovich Bose field.\cite{Burdin2002}

\begin{figure}[hbt]
\includegraphics*[width=0.46\textwidth,angle=0]{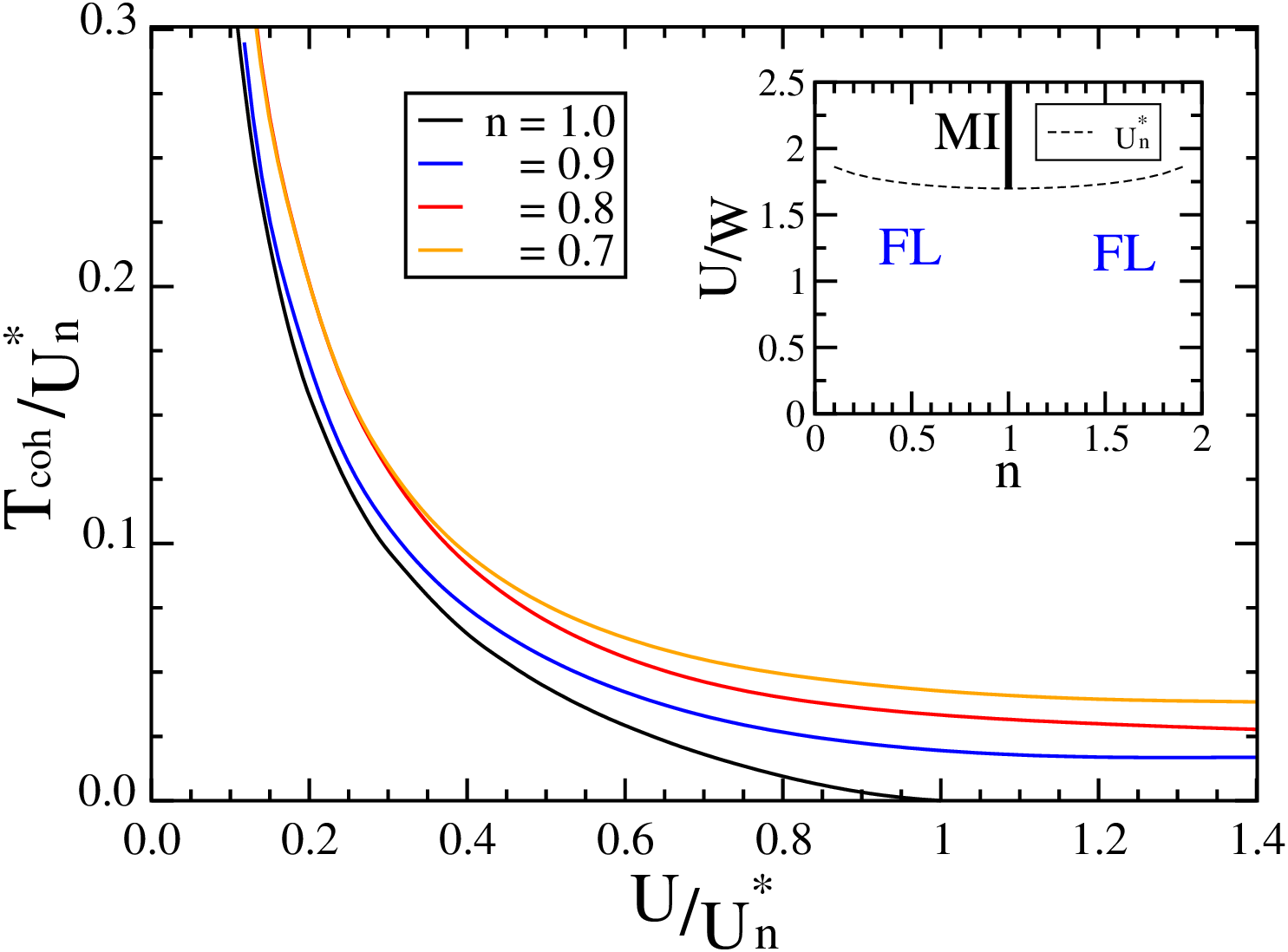}
\protect\caption{Coherence temperature $T_\textrm{coh}$ as a function of interaction strength $U$ for different fillings, using the semicircular DOS (see text).
Away from half-filling, the system is metallic, and there is no transition to a Mott insulator at $T=0$.
Due to the particle-hole symmetry of the model, the $T_\textrm{coh}$ curves for $n=0.9$, $0.8$ and $0.7$ are equivalent to those for $n=1.1$, $1.2$ and $1.3$, respectively. Away from $n=1$ we calculate the temperature $T_\textrm{coh}$ as the minimum of $\frac{\partial Z}{\partial T}$ (cf. Fig. \ref{n-080_Z}).
Here, $T_\textrm{coh}$ defines a crossover between Fermi liquid and bad metal regimes.
 $T$ and $U$ are rescaled by the correlation scale $U^*_n$ (see equation \eqref{Ustar}).
{\bf Inset}: $T=0$ phase diagram of the Hubbard model in the $U-n$ plane for the semicircular DOS case, using its bandwidth $W$ as the energy scale. At half-filling ($n=1$) and for $U>U_c$ the system is a Mott insulator, otherwise it is a Fermi liquid. The dashed line shows the behaviour with $n$ of the zero temperature correlation scale $U^*_n$ measured in units of $W$.
\label{Tcoh-generic}}
\end{figure}

In this paper, we revisit the saddle point free-energy functional in the one band Hubbard model, as originally presented by Kotliar and Ruckenstein (equation 6 of Ref. [\onlinecite{Kotliar1986}]), and investigate its properties for finite temperature and generic band filling.
Using several band structures we unify the results by a proper scaling of the different cases and compare with DMFT.
Our results shows that the dependence of the slave-boson results on the band structure details manifests itself only in the value of a zero temperature energy scale, and hence a universal behaviour is found. Also, it
gives a good qualitative agreement with numerical results in a wide range of parameters and allows a simple physical description of the correlation effects at finite temperature.
In Fig. \ref{n-100_Tcoh} we summarise our findings with the phase diagram for the one band Hubbard model at half-filling ($n=1$).
Fig. \ref{Tcoh-generic} shows results for a doped Mott insulator ($n<1$) for the case of a semicircular density of states (DOS).
There are two main results to emphasize here.
First, as we approach the Mott MIT the coherence temperature $T_\textrm{coh}$ is orders of magnitude lower than a proper correlation scale, cf. equation \eqref{Ustar}, which is of the order of the DOS bandwidth $W$.
This is consistent with what is observed in a wide range of strongly correlated electron materials, i.e., $T_\textrm{coh}$ is much lower than the scale of the non-interacting Fermi gas.
The second important result is the universal behaviour that emerges after a rescaling of temperature and interaction with the scale of equation \eqref{Ustar}.

The organization of the paper is as follows: In Section \ref{method} we briefly review the Kotliar-Ruckenstein (KR) slave-boson method for the single-band Hubbard model.
In Section \ref{results} we present our results for the temperature dependence of the quasiparticle weight, double occupancy, entropy and charge compressibility.
We link the temperature where the quasiparticle weight strongly decreases to the coherence temperature in strongly correlated materials that signal the crossover to a bad metal regime.
We find this temperature is much lower than the uncorrelated Fermi temperature scale.
Also a general agreement with DMFT is found in the dependence of all the quantities with $T$ and $U$, and a universal behaviour is found that is independent of the band structure.
The stability of the method is analysed in terms of the positivity of the charge compressibility.
Finally, concluding remarks and future directions are discussed in Section \ref{conclusion}.

\section{Model and method}
\label{method}

Our starting point is the one band Hubbard model for interacting electrons on a lattice:
\begin{equation}
\hat{H}=\sum_{i,j,\sigma} t_{ij} \hat{c}^\dag_{i\, \sigma} \hat{c}_{j\, \sigma}+U\sum_i \hat{n}_{i\, \uparrow} \hat{n}_{i\, \downarrow} - \mu \sum_i \hat{n}_i\ . \label{hubbard}
\end{equation}
As usual, $\hat{c}^\dag_{i\, \sigma}\ ( \hat{c}_{i\, \sigma})$ creates (anihilates) an electron with spin $\sigma$ at the site $i$, $\hat{n}_{i\, \sigma} = \hat{c}^\dag_{i\, \sigma}\ \hat{c}_{i\, \sigma}$ is the corresponding occupation number, and $\hat{n}_i = \hat{n}_{i\, \uparrow} + \hat{n}_{i\, \downarrow}$.
The hopping matrix element is $t_{ij}$, $U$ is the Coulomb on-site repulsion, and $\mu$ the chemical potential that fixes the average electronic density $n$.

We use the KR slave bosons as in their original work.\cite{Kotliar1986}
The KR method, in order to take into account the correlation effects, map the original fermionic local configurations to a mixed fermionic-bosonic model with local constraints.
The four slave-boson operators $\hat{e}_i$, $\hat{p}_{i\, \sigma}$ and $\hat{d}_i$, denote the empty, singly occupied, and doubly occupied states, respectively.
The corresponding occupation numbers $\hat{e}^\dag_i\, \hat{e}_i$, $\hat{p}^\dag_{i\, \sigma}\, \hat{p}_{i\, \sigma}$ and $\hat{d}^\dag_i\, \hat{d}_i$ represent the projector on the four possible states on the site $i$: $\vert 0 \rangle$, $\vert \uparrow \rangle$, $\vert \downarrow \rangle$ and $\vert \uparrow\downarrow \rangle$.
The physical electron operator is represented as $\hat{c}_{i\, \sigma}=\hat{z}_{i\, \sigma} \hat{f}_{i\, \sigma}$, where $\hat{f}_{i\, \sigma}$ is a fermionic operator and
\begin{align}
\hat{z}_{i\, \sigma} = & \frac{1}{\sqrt{1-\hat{d}^\dag_i\, \hat{d}_i-\hat{p}^\dag_{i\, \sigma}\, \hat{p}_{i\, \sigma}}}\, (\hat{e}^\dag_i\, \hat{p}_{i\, \sigma} + \hat{p}^\dag_{i\, \bar{\sigma}} \hat{d}_i) \nonumber \\
& \frac{1}{\sqrt{1-\hat{e}^\dag_i\, \hat{e}_i-\hat{p}^\dag_{i\, \bar{\sigma}}\, \hat{p}_{i\, \bar{\sigma}}}} \label{zdefinition}\ .
\end{align}
The introduction of the bosonic degrees of freedom leads to an enlarged Hilbert space and, in order to maintain the physical Hilbert space, we need to restrict the auxiliary operators with the following three constraints:
\begin{equation}
\hat{e}^\dag_i\, \hat{e}_i+ 
\hat{p}^\dag_{i \uparrow}\, \hat{p}_{i \uparrow} +
\hat{p}^\dag_{i \downarrow}\, \hat{p}_{i \downarrow} +
\hat{d}^\dag_i\, \hat{d}_i = 1
\end{equation}
and
\begin{equation}
\hat{f}^\dag_{i\, \sigma}\, \hat{f}_{i\, \sigma}  =
\hat{p}^\dag_{i\, \sigma}\, \hat{p}_{i\, \sigma} +
\hat{d}^\dag_i\, \hat{d}_i
\end{equation}
There is freedom in the choice of $\hat{z}_{i\, \sigma}$, but it ceases to be valid once we make approximations on the model.
In this work we use \eqref{zdefinition}, which was originally presented by KR, because it recovers the physics of the non-interacting $U=0$ limit in the mean-field approximation.\cite{Kotliar1986,Lavagna1991}\\
The Hubbard Hamiltonian in the new representation is,
\begin{align}
\hat{H} = & \sum_{i,j,\sigma} t_{ij} \hat{z}^\dag_{i \sigma} \hat{z}_{j \sigma} \hat{f}^\dag_{i \sigma} \hat{f}_{j \sigma}+U\sum_i \hat{d}^\dag_i \hat{d}_i - \mu \sum_i
\hat{n}^f_i
\nonumber \\
& + \sum_i \lambda^{(1)}_i  \left( \hat{e}^\dag_i\, \hat{e}_i+\hat{p}^\dag_{i \uparrow}\, \hat{p}_{i \uparrow}+\hat{p}^\dag_{i \downarrow}\, \hat{p}_{i \downarrow}+\hat{d}^\dag_i\, \hat{d}_i-1\right) \nonumber \\
& + \sum_{i\, \sigma} \lambda^{(2)}_{i\, \sigma} \left( \hat{f}^\dag_{i\, \sigma}\, \hat{f}_{i\, \sigma}-\hat{p}^\dag_{i\, \sigma}\, \hat{p}_{i\, \sigma}-\hat{d}^\dag_i\, \hat{d}_i \right)\ ,
\end{align}
where $\hat{n}^f_i = \hat{f}^\dag_{i\uparrow}\hat{f}_{i\uparrow} + \hat{f}^\dag_{i\downarrow}\hat{f}_{i\downarrow}$, and we have already added the Lagrange multipliers $\lambda^{(1)}_i$ and $\lambda^{(2)}_{i\, \sigma}$ to enforce the constraint on each site.
The mapping $\hat{n}_{i\uparrow}\hat{n}_{i\downarrow} \rightarrow \hat{d}^\dag_i \hat{d}_i$ and $\hat{n}_i \rightarrow \hat{n}^f_i$ are justified by noting that they yield the same results when applied to the four states site basis.\cite{Fresard1992,*Fresard1992b}

To this point the representation is exact, and further treatment is impossible without some approximation. We perform the saddle-point and paramagnetic approximations, as presented in Ref. [\onlinecite{Kotliar1986}], and they can be summarised in the following steps:
\begin{itemize}
\item The partition function $\mathcal{Z} = \textrm{Tr}\left( e^{-\beta\, \hat{H}} \right)$ is written as a functional integral over the fermion and boson coherent states: $e_{i\, (\tau)}$, $p_{i\, \sigma\, (\tau)}$, $d_{i\, (\tau)}$ and $f_{i\, (\tau)}$ are now complex bosonic/fermionic fields and $\tau$ is the imaginary time. Integrating out the fermionic fields we express the partition function $\mathcal{Z}$ in terms of an effective action for the bosons.
\item We carry out the saddle-point approximation over the bosonic fields and the Lagrange multipliers. The bosonic fields are replaced by their extreme values, which are assumed to be real, and they are site- and time-independent: 
$e_{i\, (\tau)} \rightarrow e$, $p_{i\, \sigma\, (\tau)} \rightarrow p_\sigma$, $d_{i\, (\tau)} \rightarrow d$, $\lambda^{(1)}_i \rightarrow \lambda^{(1)}$ and $\lambda^{(2)}_{i\, \sigma} \rightarrow \lambda^{(2)}_\sigma$.\\
The free energy per site is,
\begin{widetext}
\begin{equation}
f = U\, d^2 - \sum_\sigma  \lambda_{\sigma}^{(2)} \left( p_\sigma^2 + d^2 \right) + \lambda^{(1)} \left( e^2 + \sum_\sigma p_\sigma^2 + d^2 - 1 \right) -\frac{1}{\beta} \sum_{\sigma} \int_{-\infty}^\infty \rho (\varepsilon) \ln \left(1 + e^{-\beta\, \left[q_\sigma\, \varepsilon - \mu + \lambda_{\sigma}^{(2)}\right] } \right) \textrm{d} \varepsilon\ , \label{free-energy}
\end{equation}
\end{widetext}
%\begin{align}
%f(T) = & \ U\, d^2 - \sum_\sigma  \lambda_{\sigma}^{(2)} \left( p_\sigma^2 + d^2 \right) + \lambda^{(1)} \left( e^2 + \sum_\sigma p_\sigma^2 + d^2 - 1 \right)\nonumber \\
% & -\frac{1}{\beta} \sum_{\sigma} \int_{-\infty}^\infty \rho (\varepsilon) \ln \left(1 + e^{-\beta\, (q_\sigma\, \varepsilon - \mu - \sigma h + \lambda_{\sigma}^{(2)}) } \right) \textrm{d} \varepsilon\ , \label{free-energy}
%\end{align}
which is equation (6) of Ref. [\onlinecite{Kotliar1986}]. Here $\rho(\varepsilon)$ is the electronic DOS, and $q_\sigma = z_\sigma^2$ is the band-renormalization factor.
\item We minimize the free energy against all the bosonic fields $e$, $p_\sigma$, $d$, $\lambda^{(1)}$ and $\lambda_{\sigma}^{(2)}$
\item We perform the paramagnetic approximation: $p_\sigma \rightarrow p$, $\lambda^{(2)}_\sigma \rightarrow \lambda^{(2)}$ and $q_\sigma \rightarrow q$.
The minimization equations, together with the paramagnetic approximation, allow expressing the problem only in terms of the double occupancy number $d^2$ and the parameters of the system $U$, $n$ and $\rho(\varepsilon)$. This leads to the relations $p^2 = n/2 - d^2$ and $e^2 = 1-n+d^2$, and finally to the following self-consistent equations,
\begin{align}
  \frac{n}{2} = & \int_{-\infty}^{+\infty} \rho(\varepsilon)\ \frac{1}{1+e^{\beta(q\, \varepsilon - \mu + \lambda^{(2)})}}\ \textrm{d} \varepsilon \label{selfn}\\
%  0 = &\, \frac{U}{2} +  q\, \bar{\varepsilon}\ \frac{p^2 - d\, e}{e\, d\, p^2} \label{selfM} \\
 \frac{U}{2} = &\, q\, \bar{\varepsilon}\, \left( \frac{1}{p^2}- \frac{1}{e\, d} \right) \label{selfM} \\
  \lambda^{(2)}\! = &\, \frac{U}{2}\! + \frac{q\, \bar{\varepsilon}}{d} \left(1- \frac{d}{e}\, \frac{1-\frac{n}{2}}{\frac{n}{2}} \right)\!\! \left( \frac{1}{d+e} + \frac{d}{1-\frac{n}{2}} \right) \label{selfL}
\end{align}
where 
\begin{equation}
\bar{\varepsilon}(T) =  \int_{-\infty}^{+\infty}\! \varepsilon\ \rho(\varepsilon)\ \frac{1}{1+e^{\beta(q \varepsilon - \mu + \lambda^{(2)})}}\ \textrm{d} \varepsilon
\label{equationebar}
\end{equation}
is the uncorrelated energy per site and spin, and
\begin{equation}
q = \frac{p^2\, \left(d+e\right)^2}{\left(1-\frac{n}{2}\right)\frac{n}{2}} \ .
\label{equationq}
\end{equation}
For simplicity, we have kept the notation for $e$ and $p$.
\end{itemize}
Equations (\ref{selfn}-\ref{selfL}), with definitions (\ref{equationebar}-\ref{equationq}), has to be solved self-consistently for the quantities $d$, $\mu$ and $\lambda^{(2)}$.
The free energy per site is now,
\begin{equation}
f(T)\! = U d^2 - \lambda^{(2)} n -\frac{2}{\beta} \int_{-\infty}^\infty\!\!\! \rho (\varepsilon) \ln \left(1 + e^{-\beta (q \varepsilon - \mu + \lambda^{(2)}) } \right)\! \textrm{d} \varepsilon\, . \label{free-energy-self}
\end{equation}

A useful rewriting of the self-consistent equation \eqref{selfM} allows us to see that the density of holes $\delta=1-n$ and the dimensionless interaction strength $U/U^*(T,\delta)$ as the relevant parameters of the problem.\cite{Fresard1992,*Fresard1992b,Dao2017} The coupling scale $U^*(T,\delta)$ is defined as
$U^*(T,\delta) \equiv \frac{-16}{1-\delta^2}\, \bar{\varepsilon}(T)$.
We note that $U^*(T,\delta)$ is a self-consistent parameter for every filling value $n$ and finite value of $T$, as it depends on the self-consistent values of $q$, $\lambda^{(2)}$ and $\mu$ through the Fermi-Dirac distribution function of definition \eqref{equationebar}. For $T=0$ it only depends on the uncorrelated DOS and the filling $n$, and we denote it
\begin{equation}
U^*_n \equiv U^*(0,1-n) =\frac{-16}{1-\delta^2}\int_{-\infty}^{\varepsilon_\mathrm{F}} \varepsilon \rho (\varepsilon) \mathrm{d}\varepsilon \label{Ustar}\ .
\end{equation}
Throughout the paper we use $U^*_n$ as a unit for $U$, $T$ and $\mu$.
For the particular case of half-filling ($n=1$) we have,
\begin{equation}
e^2=d^2\ , \ \ p^2=1/2-d^2
\end{equation}
and the self-consistent equations
\begin{align}
%    \frac{1}{2} =&\ \int_{-\infty}^{+\infty} \rho(\varepsilon)\ \frac{1}{1+e^{\beta(q \varepsilon - \mu + \lambda^{(2)})}}\ \textrm{d} \varepsilon \label{selfn-hf}\\ 
    \lambda^{(2)} =&\ \frac{U}{2} \label{selfL-hf}\\
    d^2(T) =&\ \frac{1}{4}\left( 1 - \frac{U}{U^*(T,0)}\right) \label{selfM-hf} \\
    q(T) =&\ 1 - \left( \frac{U}{U^*(T,0)}\right)^2 \label{selfq-hf}
\end{align}
reproduce the $T=0$ result of the original KR paper,\cite{Kotliar1986} with
\begin{equation}
    U_c = U^*(0,0) = U^*_1 = -16\ \bar{\varepsilon}(T=0)\label{Ucritic}
\end{equation}
the critical value of the interaction $U$ where the metal-insulator transition occurs.

As can be observed from the results discussed  in the next section, the calculated quantities are nearly insensitive to details of the DOS when energy and temperature are scaled with $U^*_n$.
We can easily understand this behaviour at $T=0$ by noting that the DOS appears only in the $n$ and $\bar{\varepsilon}$ equations.
The occupation number $n$ is a fixed parameter for the system, and different shapes of the DOS will not affect it and only change the value of the Fermi energy $\varepsilon_\mathrm{F}$.
The details of the DOS does affect the value of the uncorrelated kinetic energy $\bar{\varepsilon}$, and hence the complete independence of the $T=0$ results once renormalised with this quantity.
In a similar way, the temperature $T$ appears in the $n$ and $\bar{\varepsilon}$ expressions and so we might expect a weak dependence on the DOS  for finite $T$.
Previously, Hasegawa observed that magnetic properties such as the N\'{e}el temperature and magnetization, calculated with the temperature dependent slave-boson method, depend weakly on the details of the DOS.\cite{Hasegawa1990a}

\section{Results}
\label{results}

In this paper, we perform the calculation for several different band structures, making use of the mapping between the wave vector and the energy form of the integrals,
\begin{equation}
\int_{-\infty}^{\infty} \gamma (\varepsilon) \rho (\varepsilon) \mathrm{d}\varepsilon =
\frac{1}{N} \sum_{\bf k} \gamma (\varepsilon_{\bf k})\ 
\end{equation}
for a function $\gamma$, where $\displaystyle \varepsilon_{\bf k}=2 \sum_{{\bf R}_j-{\bf R}_i} t_{ij}\, \cos \left[ {\bf k}\cdot \left( {\bf R}_j-{\bf R}_i \right) \right] $ is the uncorrelated dispersion relation in the paramagnetic state, $N$ is the number of sites in the lattice, and the sum in $\bf k$ is over the first Brillouin zone of the reciprocal lattice for a real space Bravais lattice with points $\lbrace {\bf R}_i \rbrace$.
We use the wave vector formulation for square\cite{squarenote}
 and triangular lattices, up to $40,000$ and $43,200$ sites, respectively. Also, as DOS we use $\rho(\varepsilon) = \frac{8}{\pi} \frac{1}{W^2} \sqrt{\left(\frac{W}{2}\right)^2-\varepsilon^2}$ (semicircular) for the Bethe lattice in the infinite connectivity limit,\cite{Georges1996} $\rho(\varepsilon) = \frac{1}{W}$ as a flat DOS, where $W$ is the bandwidth, and $\rho(\varepsilon) = \frac{1}{\sqrt{\pi}\, t^*} \mathrm{exp} \left( -\varepsilon^2/{t^*}^2 \right)$ for the hypercubic lattice in the limit of infinite dimension, with $t^*$ the scaling of the hopping.\cite{Metzner1989a}

Except when comparing the results for different band structures, we present our results for the case of the semicircular DOS. It allows a quantitative comparison with DMFT results since this method becomes exact in the infinite-dimensional limit of the Bethe lattice.\cite{Georges1996}

\subsection{Quasiparticle weight and coherence temperature}
The quasiparticle weight $Z$ is the spectral weight of the quasiparticle peak at the Fermi energy.
This peak involves the coherent excitations that form the Fermi liquid, and its spectral weight decreases as the metal-insulator transition is approached.
It can also be understood as the renormalization factor of the fermionic band, which is $q$ in this formulation.
For a Fermi liquid it implies that, as the specific heat is linear in $T$ at low temperatures, the slope is $1/Z$ times larger than the non-interacting value.
\begin{figure}[hbt]
\includegraphics*[width=0.46\textwidth,angle=0]{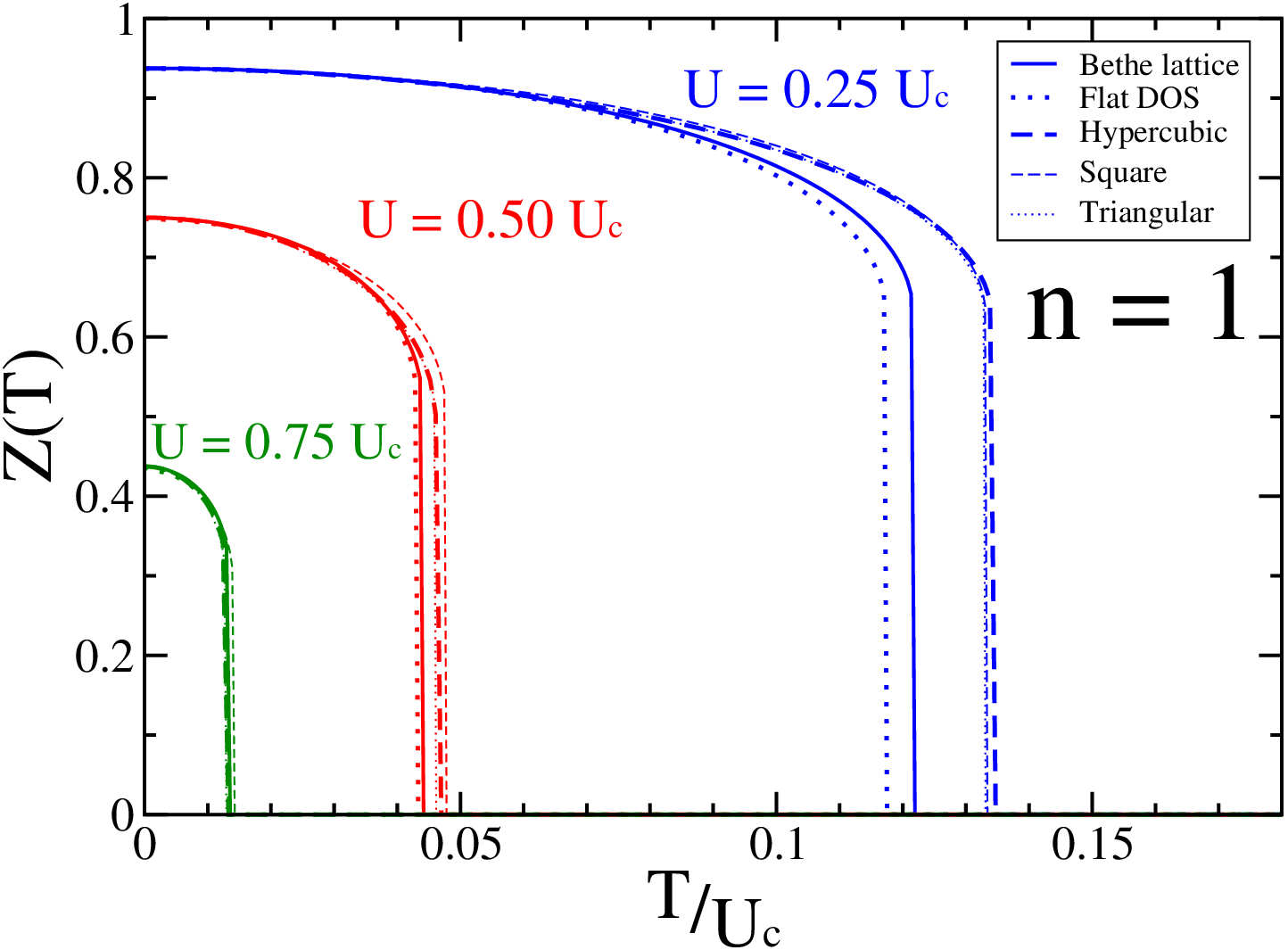}
\protect\caption{Fermi liquid quasiparticle weight $Z$, at half-filling, as a function of temperature, for several values of $U/U_c$ and band structures. Note the weak dependence on the shape of the DOS.
An increment of the temperature $T$ or the interaction $U$ diminishes the quasiparticle weight $Z$.
We identify the emergence of the bad metal with the collapse of $Z$ at the coherence temperature $T_\textrm{coh}$.
The dependence of $T_\textrm{coh}$ on $U$ is shown in Fig. \ref{n-100_Tcoh}.
\label{n-100_Z}}
\end{figure}

In Fig. \ref{n-100_Z} we  show, for several different band structures, the temperature dependence of the quasiparticle weight $Z$ at half-filling and several values of the interaction $U$, being $T$ and $U$ properly scaled.
It decreases with increasing $T$ and jumps to zero at $T_\textrm{coh}$, the coherence temperature of the fermionic quasiparticle. As expected, its behaviour does not change significantly between different band structures for $T>0$, and does not change at all for $T=0$.
Fig. \ref{n-100_Tcoh} shows the dependence of $T_\textrm{coh}$ on the interaction strength (color points), and the corresponding phase diagram.
The black dashed line is the approximate expression obtained in Ref. [\onlinecite{Fresard1997}], i.e.,
\begin{equation}
\label{FKequation}
\frac{T_\textrm{coh}}{U_c} \simeq \frac{\big[ \, 1 - \frac{U}{U_c}\, \big]^2}{8\, \ln(2)}\ .
\end{equation}
We also plot the renormalized Fermi temperature scale $T^*_\textrm{F} \equiv Z(T\!=\!0)\, T^{0}_\textrm{F} = Z(T\!=\!0)\, \frac{W}{2}$, obtained by considering the system as a strongly renormalized Fermi liquid\cite{Georges1996} (blue dashed line in Fig. \ref{n-100_Tcoh}), where $T^0_\textrm{F} = \varepsilon^0_\textrm{F}$ is the Fermi temperature for $U=0$.
From this figure we conclude that the slave-boson method reproduces the important property that strongly correlated materials exhibit a very low coherence temperature close to a Mott MIT.
For interaction values close to the MIT, the dependence of $T_\textrm{coh}$ closely follows equation \eqref{FKequation} and is about ten times lower than the renormalized scale $T^*_\textrm{F}$, as is found with DMFT.\cite{Vucicevic2013}
Fig. \ref{n-100_d2} of the Appendix shows the behaviour of the double occupancy for the same parameters and is compared with DMFT simulations. The decrease in double occupancy with increasing temperature reveals the tendency of the system to a higher degree of localisation.

It is important to note that the slave-boson method studied here is not valid in the half-filling case for $T>T_\textrm{coh}$, ruling out any attempt to treat the high-temperature limit.
For some fixed $U<U_c$ the quantity $U^*(T,0)$ is higher than $U$ for low temperatures and decreases with increasing $T$. At $T_\textrm{coh}$ the system gets stuck in the trivial solution $d^2=0$, which was ruled out in the derivation of the self-consistent equations. The latter solution physically means a system in which each electron freezes in the site and no double occupation is allowed, and strictly speaking, the slave-boson method predicts a transition to a Mott insulator at $T=T_\textrm{coh}$.

\begin{figure}[hbt]
\includegraphics*[width=0.46\textwidth,angle=0]{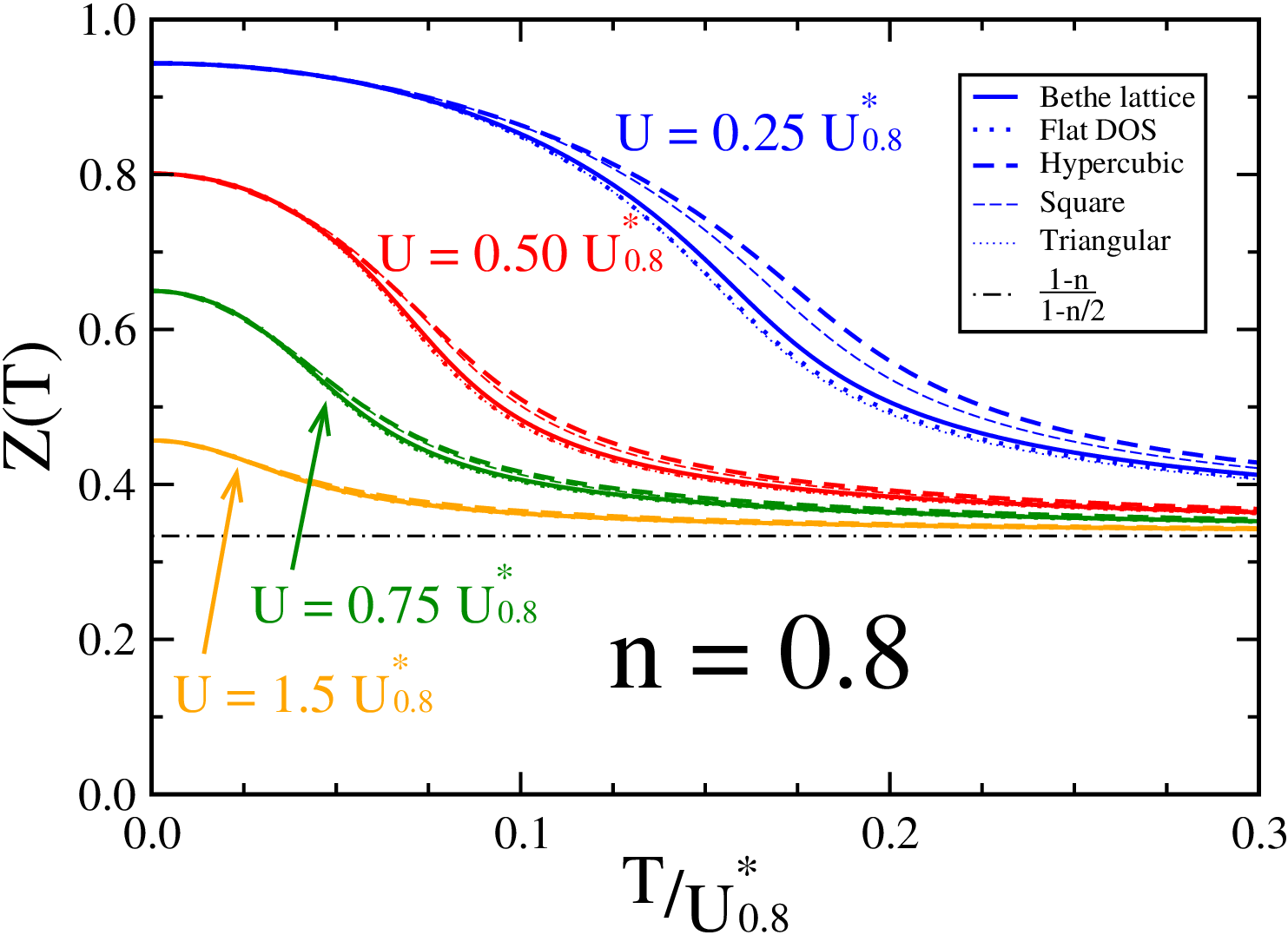}
\protect\caption{Same as Fig. \ref{n-100_Z} for $n=0.8$, with $U^*_\textrm{0.8}$ as energy unit. Values of $U/U^*_n$ larger than $1$ can be achieved because the system always remains metallic and does not go through a MIT when $U$ increase.
Even though the double occupancy goes to zero for larger temperatures (cf. Fig. \ref{n-080_d2} in the Appendix), there are still holes that allow electronic movement and the quasiparticle weight $Z$ remains finite, with a lower bound of $Z=\frac{1-n}{1-\frac{n}{2}}$.
The inflexion points in the curves for the semicircular DOS are used to calculate the coherence temperature $T_\textrm{coh}$ in Fig. \ref{Tcoh-generic}.
\label{n-080_Z}}
\end{figure}

For filling values different from $n=1$ (see Fig. \ref{n-080_Z} for $n=0.8$) the system is always metallic.
With increasing temperature the double occupancy $d^2$ goes continuously to zero (cf. Fig. \ref{n-080_d2} in the Appendix) and, from equation \eqref{equationq}, the quasiparticle weight goes to its minimum value $Z(T\rightarrow\infty)=\frac{1-n}{1-\frac{n}{2}}$.
Although the quasiparticle weight does not completely dissappear, we define the coherence temperature as the temperature at which the decrease in $Z$ is more pronounced, $\left. \frac{\partial^2 Z}{\partial T^2}\right\vert_{T_\textrm{coh}}=0$ (i.e., an inflexion point).
In a similar vein the inflexion point of the spectral density at the Fermi level with respect to the temperature has been used as one definition of a crossover line between the Fermi liquid and the bad metal regime.\cite{Vucicevic2013}
Similarly to the previous discussion for the half-filling case, we can assume that the slave-boson method ceases to be valid at temperatures higher than $T_\textrm{coh}$, where the change with temperature towards the trivial solution $d^2=0$ is more pronounced. This assumption is later justified by noting that the method is unstable for $T>T_\textrm{coh}$ for relatively small doping (see Section \ref{compressibility} and Fig. \ref{n-100_chi}).

In Fig. \ref{Tcoh-generic} we show the dependence of $T_\textrm{coh}$ vs $U$ for different values of filling $n$ using the DOS of the Bethe lattice. Although $T_\textrm{coh}$ is modified, it still remains very low when the interaction $U$ is of the order of the correlation scale $U^*_n$ and larger. The inset of Fig. \ref{Tcoh-generic} shows the $T=0$ phase diagram of the Hubbard model in the $U-n$ plane, and the dependence of the scale $U^*_n$ with $n$ (dashed line).

\subsection{Entropy}
\label{entropy}

We calculate the entropy density per site through the thermodynamic relation
\begin{equation}
s(T) = \beta\ \left(\overline{u} - f \right)% + n\, \frac{\partial \mu'}{\partial T}\ ,
\end{equation}
where $f$ is the free energy of equation \eqref{free-energy-self}, and,
\begin{equation}
\overline{u}(T) = U\, d^2 + 2\, q\, \bar{\varepsilon} - \mu\, n
\end{equation}
is the (correlated) energy per lattice site. 
\begin{figure}[hbt]
\includegraphics*[width=0.46\textwidth,angle=0]{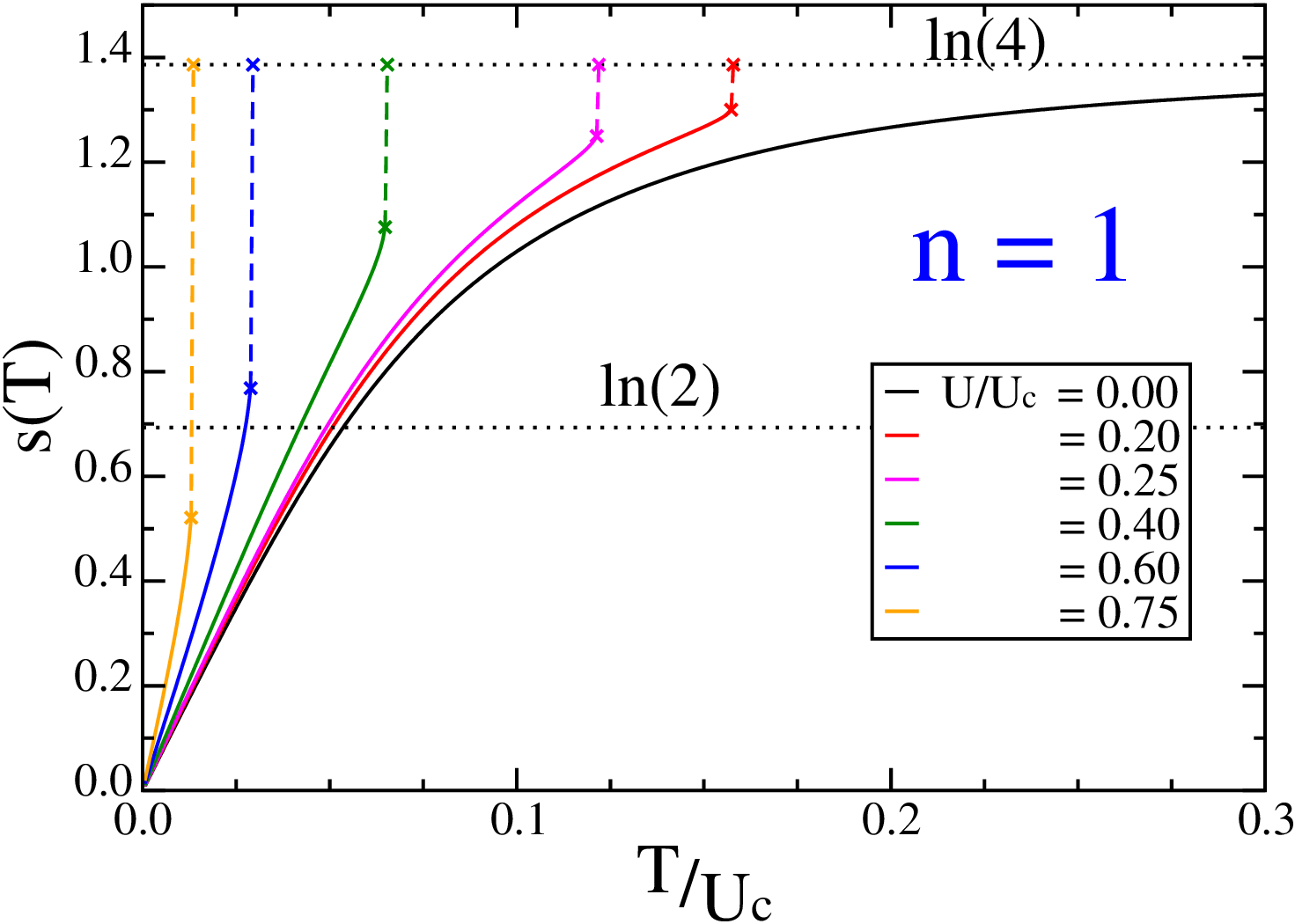}
\protect\caption{Entropy density at half-filling as a function of temperature for different interaction values $U/U_c$, using the semicircular DOS.
The $\ln (4)$ dotted line correspond to the $d^2=0$ limit and to a regime with local charge and spin fluctuations; and the $\ln (2)$ value corresponds to localised non-interacting spins.
The increase of entropy with $U$, accompanied by the decrease of $d^2$ with $T$, is an analog of the Pomeranchuk effect in liquid $^3$He and is consistent with DMFT results in the same model.\cite{Werner2005}
For a comparison with DMFT results in the hypercubic lattice see Fig. \ref{n-100_S-hyper} in the Appendix.
\label{n-100_S}}
\end{figure}

In Fig. \ref{n-100_S} we display the entropy density as a function of the temperature for various $U/U_c$ values, for the half-filling case.
At low temperatures the entropy increases linearly with temperature, as expected for a Fermi liquid.
We also show with dotted lines the corresponding values of free on-site spin fluctuations, $s=\ln (2)$, and the high-temperature limit, $s=\ln (4)$, where charge and spin fluctuations are completely free.
Similar to DMFT results in the low-temperature Fermi liquid type regime,\cite{Pakhira2015b} we obtain that the entropy increases as the system becomes further correlated.
As we increase $U$ and approach the Mott phase, the value of the entropy at $T_\textrm{coh}$ (end crosses at each solid line) decrease.
The latter can be related to the appearance of kink-like features of the entropy found with DMFT\cite{Pakhira2015b,Toschi2007} (see Fig. \ref{n-100_S-hyper} in the Appendix for a comparison with DMFT results).

The higher degree of localisation as temperature increases, expressed in the decreasing of $d^2$ (cf. Fig. \ref{n-100_d2}), is characteristic of a strongly correlated Fermi liquid in a regime dominated by spin fluctuations.
This is a direct analog of the Pomeranchuck effect in liquid $^3$He: because the spin entropy is much larger in a localized state than in the Fermi liquid, for increasing temperature the system can lower the free energy $f=\overline{\varepsilon}-T s$ by an increase of the localisation of the particles\cite{Georges1992a,*Georges1993,Werner2005} (one can go from the itinerant to the localized phase upon heating).
The relation between $d^2(T)$ and $s(T)$ can also be understood from the Maxwell thermodynamic relation $\Big( \frac{\partial s}{\partial U}\Big)_T = - \Big( \frac{\partial d^2}{\partial T}\Big)_U$.%, that arise from the dependence of the free energy with the double occupancy.\cite{Werner2005}

\subsection{Charge compressibility}
\label{compressibility}

We numerically evaluated the charge compressibility
\begin{equation}
\chi_c \equiv \frac{\partial n}{\partial \mu}\ .
\end{equation}
This is a useful quantity in the study of the metal-insulator transition because it measures the particle stiffness of the system with a change in chemical potential.
Its inverse can be interpreted as the energy required to add or remove a particle from the system. At half-filling this agrees with the idea of the system becoming more incompressible ($\chi_c$ going to zero) as we approach the Mott insulating phase.\cite{Pakhira2015}
It is worth noting that an analytic calculation of $\chi_c$ from equation \eqref{selfn} is complicated because $n$ has a dependence on the double occupancy $d$ (due to the $d$-dependence of $q$ and $\lambda^{(2)}$), and the derivative against $\mu$ of the latter turns out to be difficult to calculate from equation \eqref{selfM}.
% This occurs because an explicit self-consistent equation for the double occupancy $d=g(d)$
\begin{figure}[hbt]
\includegraphics*[width=0.46\textwidth,angle=0]{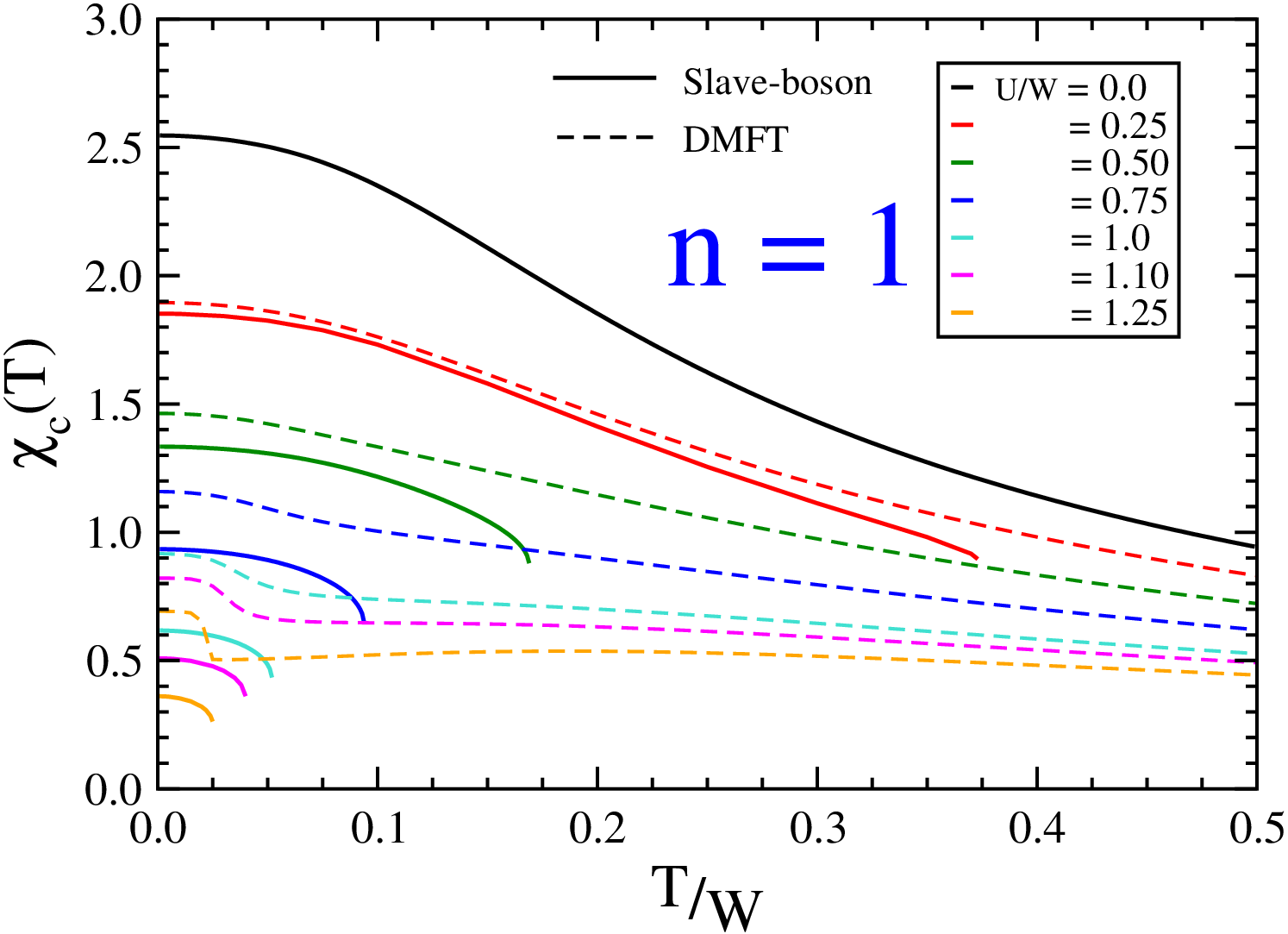}
\protect\caption{Charge compressibility $\chi_c$ as a function of temperature for the semicircular DOS, using slave-boson (solid line) and DMFT (dashed line) calculations from Ref. [\onlinecite{Pakhira2015}].
The bandwidth $W$ is used as an energy scale in order to make the comparison easier.
The decrease of charge compressibility with increasing interaction $U$ and temperature is consistent with DMFT results\cite{Pakhira2015} and with finite temperature Lanczos methods for the triangular lattice.\cite{Kokalj2013}
Also, we qualitatively reproduce the apparent kink-like feature found with DMFT at about $T\simeq 0.025\, W$ for $U = 1.25\, W$.\cite{Pakhira2015}
\label{n-100_chi}}
\end{figure}

In Fig. \ref{n-100_chi} we compare slave-boson and DMFT results\cite{Pakhira2015} for the temperature dependence of the charge compressibility for several values of $U$ at half-filling and using the semicircular DOS.
To make the comparison easier, we use the bandwidth $W$ as an energy scale, where $U_c \simeq 1.7\, W$ for this DOS.
A good agreement with the DMFT results is found, especially for low $U/W$ values. As a general feature, the charge compressibility is strongly suppressed with increasing $U$ from its noninteracting electron value, and it decreases with increasing temperature, which is also found using the finite-temperature Lanczos method for the triangular lattice\cite{Kokalj2013}.
For the interaction value $U=1.25\, W$ our result reproduces the apparent kink-like feature at $T\simeq 0.025\, W$ obtained with DMFT.\cite{Pakhira2015}
Also, for the larger $U$ values our values for $T_\textrm{coh}$, signalled in this figure by the end of the solid lines, seems to correspond to an inflexion point in the DMFT results.

\subsection{Doped Mott insulator}

\begin{figure}[hbt]
\includegraphics*[width=0.46\textwidth,angle=0]{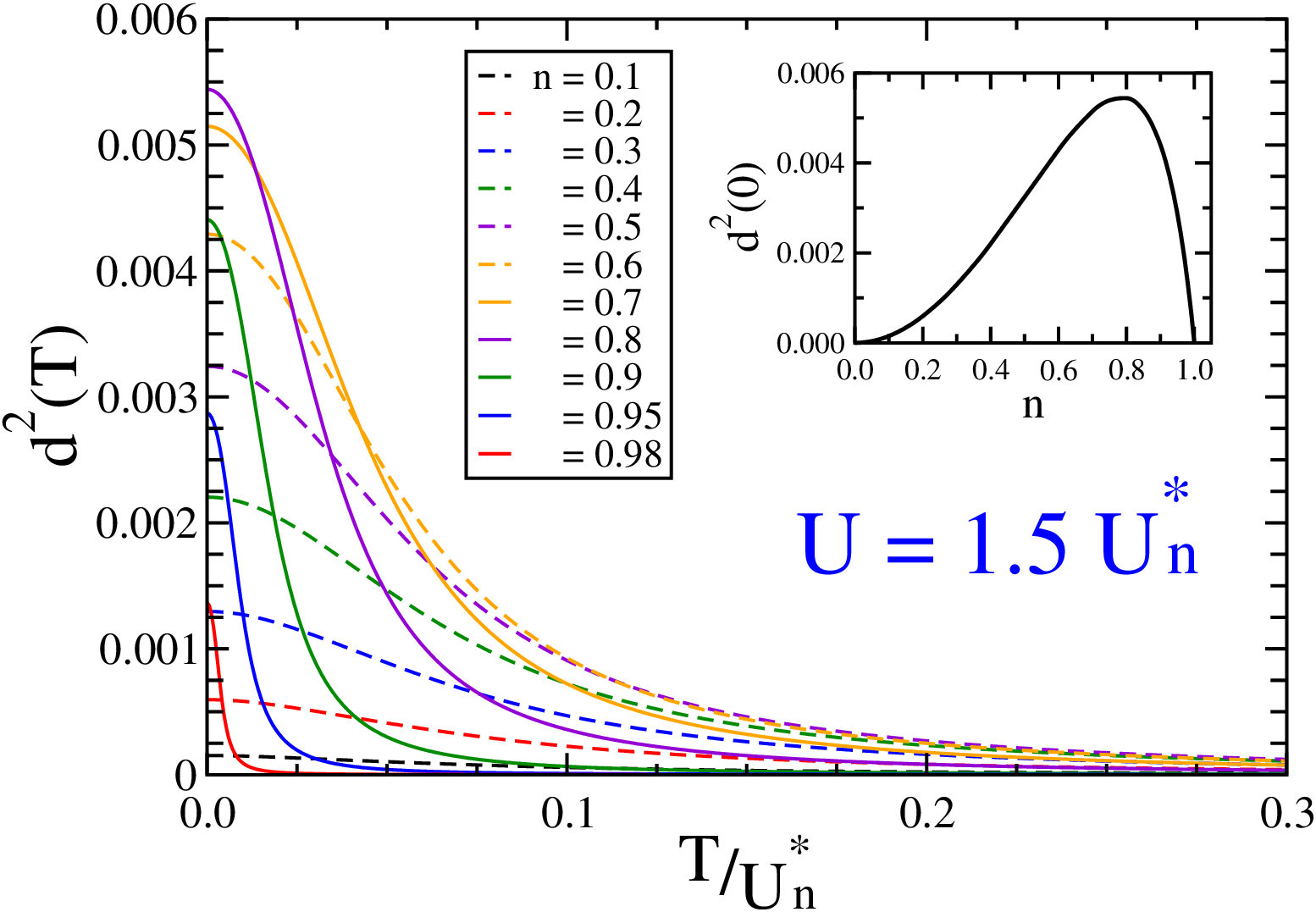}
\protect\caption{Double occupancy at $U=1.5\, U^*_n$ as a function of temperature, for several fillings.
{\bf Inset}: Dependence of $T=0$ double occupancy with the filling $n$.
As we approach the Mott phase at $n=1$ effective interactions become stronger and $d^2$ decreases towards zero.
\label{U-150_d2}}
\end{figure}

Now we focus on the behaviour as we vary the hole doping from the Mott insulator.
In Fig. \ref{U-150_d2} we plot the temperature dependence of the double occupancy $d^2$ at the electron interaction value $U= 1.5\, U^*_n$ for different fillings $n$.
As shown in Fig. \ref{Tcoh-generic}, the system remains in a metallic phase as long as $n\neq 1$ and $T < T_\textrm{coh}$. Namely, there is no critical $U_c$ away from half-filling.
The inset shows the zero temperature value of $d^2$ as a function of the electron filling.
As we approach the Mott insulator, the increasing correlation tends to localise the particles and $d^2 \rightarrow 0$ as $n \rightarrow 1$. Moving away from half-filling allows an energy gain due to hole delocalisation that is greater than the energy loss due to Coulomb repulsion $U$. By increasing the doping further, we recover a combinatorial dependence of $d^2$, but about ten times lower than the uncorrelated value $\frac{n^2}{4}$. That is, even in a highly doped Mott insulator the electronic correlations strongly suppress the double occupancy.

\begin{figure}[hbt]
\includegraphics*[width=0.46\textwidth,angle=0]{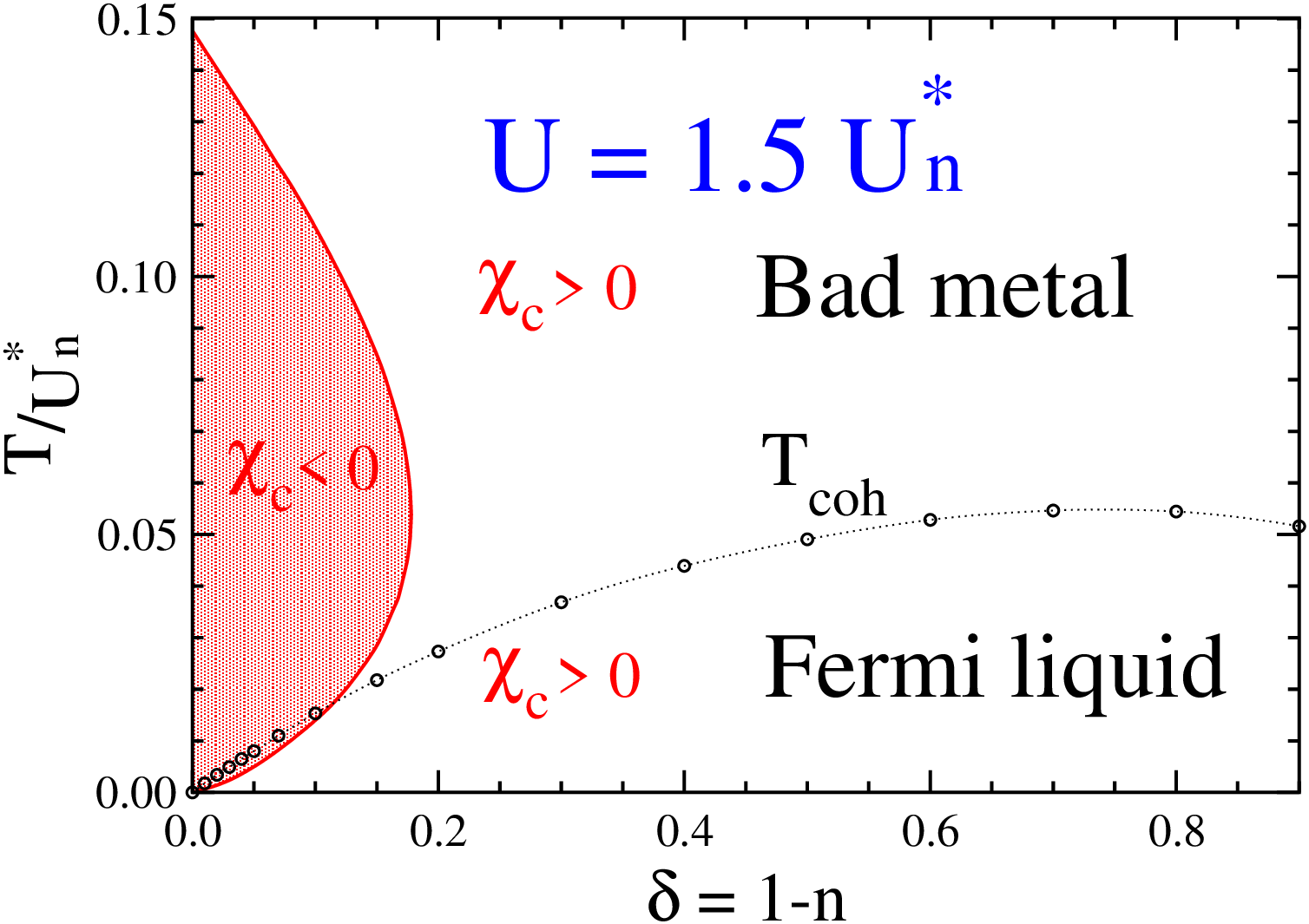}
\protect\caption{Coherence temperature $T_\textrm{coh}$ (black circles) as a function of hole doping $\delta = 1-n$, for $U=1.5\, U^*_n$ using the semicircular DOS. Near half-filling $T_\textrm{coh}$ is proportional to the doping level $T_\textrm{coh} \simeq 0.18\, \delta\, U^*_n \simeq 0.3\, \delta\, W$. This behavior is similar to the one obtained with DMFT.\cite{Deng2013a}
The red curve shows the boundary of the region at which the system becomes unstable (shaded area) due to a negative charge compressibility $\chi_c$. This region with $\chi_c<0$ does not change significantly as we vary the value of $U$ or band structure.
\label{U-150_Tcoh}}
\end{figure}
In Fig. \ref{U-150_Tcoh} we show the temperature vs. doping phase diagram for $U=1.5\, U^*_n$ with a semicircular DOS.
We obtain for low doping a linear dependence with $\delta$ of the form $T_\textrm{coh} \simeq 0.18\, \delta\, U^*_n \simeq 0.3\, \delta\, W$, which is qualitatively similar to previous DMFT results.\cite{Deng2013a,Pruschke1995}
We also obtain the shaded red region in which the charge compressibility takes negative values, which is a signature of instabilities towards phase separation, incommensurate magnetic order, or other exotic electronic phases.\cite{DeMedici2016b} For low doping, this region appears at temperatures $T\sim T_\textrm{coh}$ and higher, and this allows us to conclude that the slave-boson method remains stable in a broad region of the $T\!-\!n$ phase diagram.
A similar shape of this region is found for other values of the interaction $U$, following the general behaviour depicted in Fig. (4) of Ref. [\onlinecite{Dao2017}].
The same behaviour is found using the other band structures. This last result is not trivial, given that $\mu(T) - \mu(0) \propto \rho'(\varepsilon_\textrm{F})\, T^2$ at low temperature.\cite{Ashcroft1976} Fig. \ref{U-150_mu} in the Appendix shows $\mu$ vs. $n$ at two different temperatures for this value of $U$, where the independence with band structures is appreciated.
A recent slave-spin mean-field study of a multi-orbital Hubbard model found that the Hund's coupling is essential in the development of an instability region in the $U\!-\!\delta$ phase diagram.\cite{DeMedici2016b} This instability region departs from the Mott transition at half filling and is proposed to be related at finite-$T$ to a ``spin-freezing cross-over'', signaled by a quick decrease of $Z$, increase of inter-orbital spin-spin correlation, and suppression of inter-orbital charge-charge correlations.

\section{Concluding remarks}
\label{conclusion}

In summary, we have studied the finite temperature KR slave-boson mean-field approach of the single-band Hubbard model, and identify the temperature at which the Fermi liquid collapse with the coherence temperature found in several strongly correlated materials that signal the appearance of the bad metal regime.
For $T < T_\textrm{coh}$ our results agree with the physical picture of a renormalized Fermi liquid state, and they are in good qualitative correspondence with DMFT and finite temperature Lanczos calculations.
In particular, we find that near the Mott transition the coherence temperature is much lower than the Fermi temperature for $U=0$, i.e. $T_\textrm{coh} \ll T_\textrm{F}^0$, and in good agreement with the analytic approximation from Ref. [\onlinecite{Fresard1997}].
Also, our results shows a universal behaviour when temperature, interaction and chemical potential are scaled with a proper energy scale, making results independent of the details of the band structure used.

\section{Possible future directions}
Iron based superconductors have led to increased interest in the role of orbital degeneracy, Hund’s rule interaction $J$, and multiple bands in strongly correlated electron materials.
It was recently shown that $J$ has a conflicting effect on correlations.\cite{Georges2013a}
On the one hand, $J$ increases the critical $U$ above which a Mott insulator is formed.
On the other hand, $J$ reduces the Fermi liquid coherence temperature significantly, leading to at higher temperatures what is referred to as a ``Hund’s metal'',\cite{Haule2009a,Ong2012} which may be characterised by particularly slow spin dynamics.\cite{Werner2008,Hansmann2010}
This is the multi-band analog of the bad metal in a single band system. In a future study we plan to investigate how $T_\textrm{coh}$ varies with $J$ in multi-band models using a finite-temperature version of rotationally invariant slave-bosons\cite{Li1989,Fresard1992,*Fresard1992b,Lechermann2007,Bunemann2011,Lanata2016a,Fresard1997} or slave spins.\cite{DeMedici2016a,DeMedici2016b}
At zero temperature the latter reproduces KR slave-boson mean-field results for the single-band Hubbard model, and recently has been used to model Fe-based superconductors.\cite{DeMedici2016b}

Gaussian fluctuations from the saddle point paramagnetic solution allow to calculate the charge fluctuation matrix and give an approximate description of the upper Hubbard band.\cite{Zimmermann1997,Dao2017} On the other hand, a new Gutzwiller variational wave function involving “ghost” orbitals gives results in very good agreement with DMFT.\cite{Lanata2017}
Hopefully, a finite-temperature version of this method could also be used to investigate the emergence of the bad metal with increasing temperature.

\begin{acknowledgments}
We thank A. Camjayi, V. Dobrosavljevi{\'{c}}, R. Fr{\'{e}}sard, N. Lanat\`a, L. O. Manuel, J. Merino,  H. L. Nourse and B. J. Powell for discussions, and N. Pakhira for making available the DMFT data in Figs. \ref{n-100_chi} and \ref{n-100_S-hyper}.
This work was supported by an Australian Research Council Discovery Project, Grant No. DP160102425.
\end{acknowledgments}

\appendix*

\section{}
\label{apendix}
\begin{figure}[hbt]
\includegraphics*[width=0.46\textwidth,angle=0]{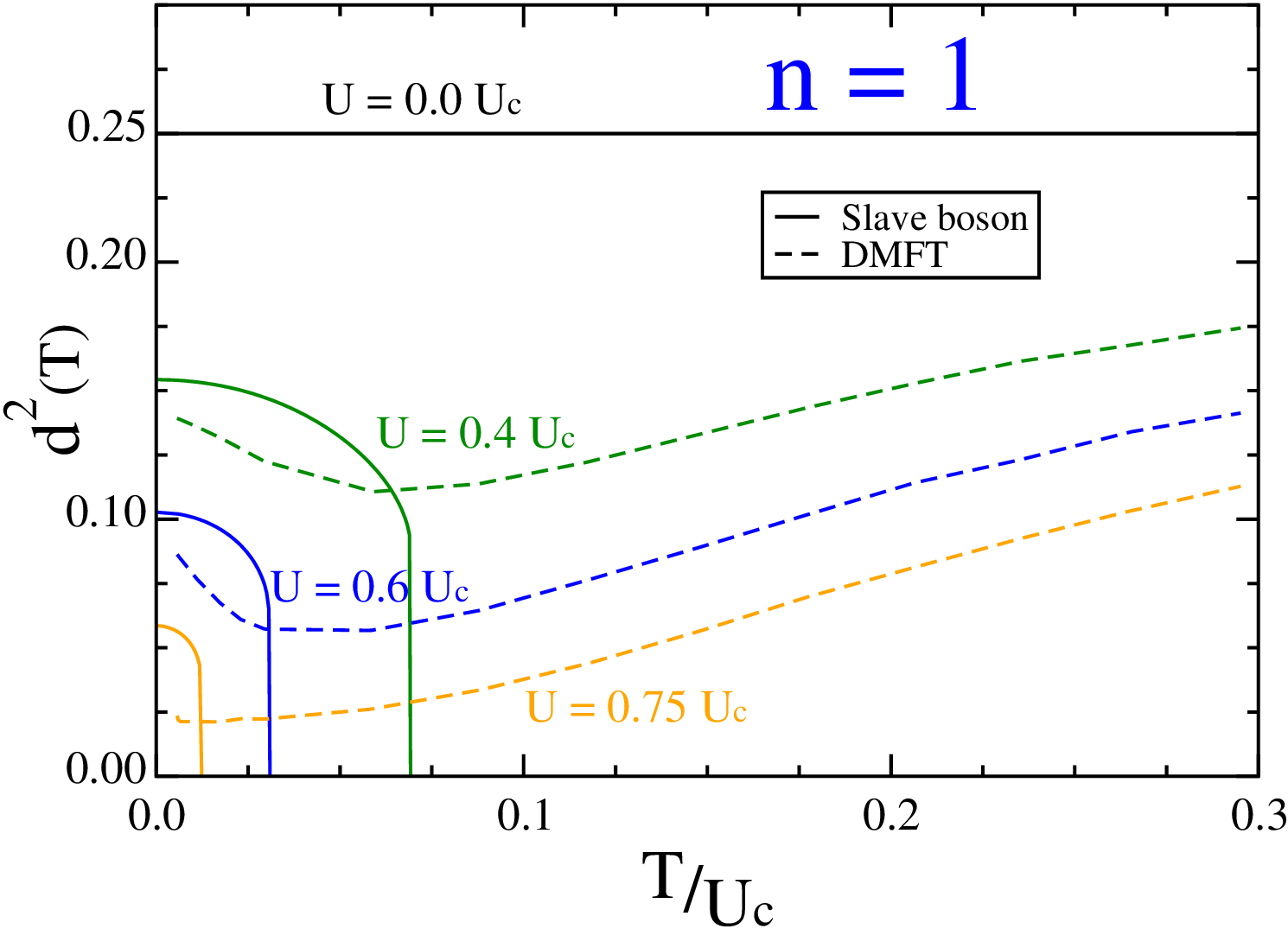}
\protect\caption{Double occupancy $d^2$ as a function of $T$ for the semicircular DOS, using slave-boson (solid line) and DMFT (dashed line) calculations using TRIQS. The decrease of $d^2$ with increasing $T$ indicates a higher degree of localisation. The slave-boson method ceases to be valid for $T>T_\textrm{coh}$, where the self-consistent equations (\ref{selfL-hf}-\ref{selfq-hf}) collapse to the trivial solution $d^2=0$. The obtained $T_\textrm{coh}$ compares favourably with the minima in the DMFT results.
\label{n-100_d2}}
\end{figure}

Fig. \ref{n-100_d2} shows the temperature dependence of the double occupancy determined by the slave-boson method (solid lines) and DMFT (dashed lines) 
\begin{figure}[hbt]
\includegraphics*[width=0.46\textwidth,angle=0]{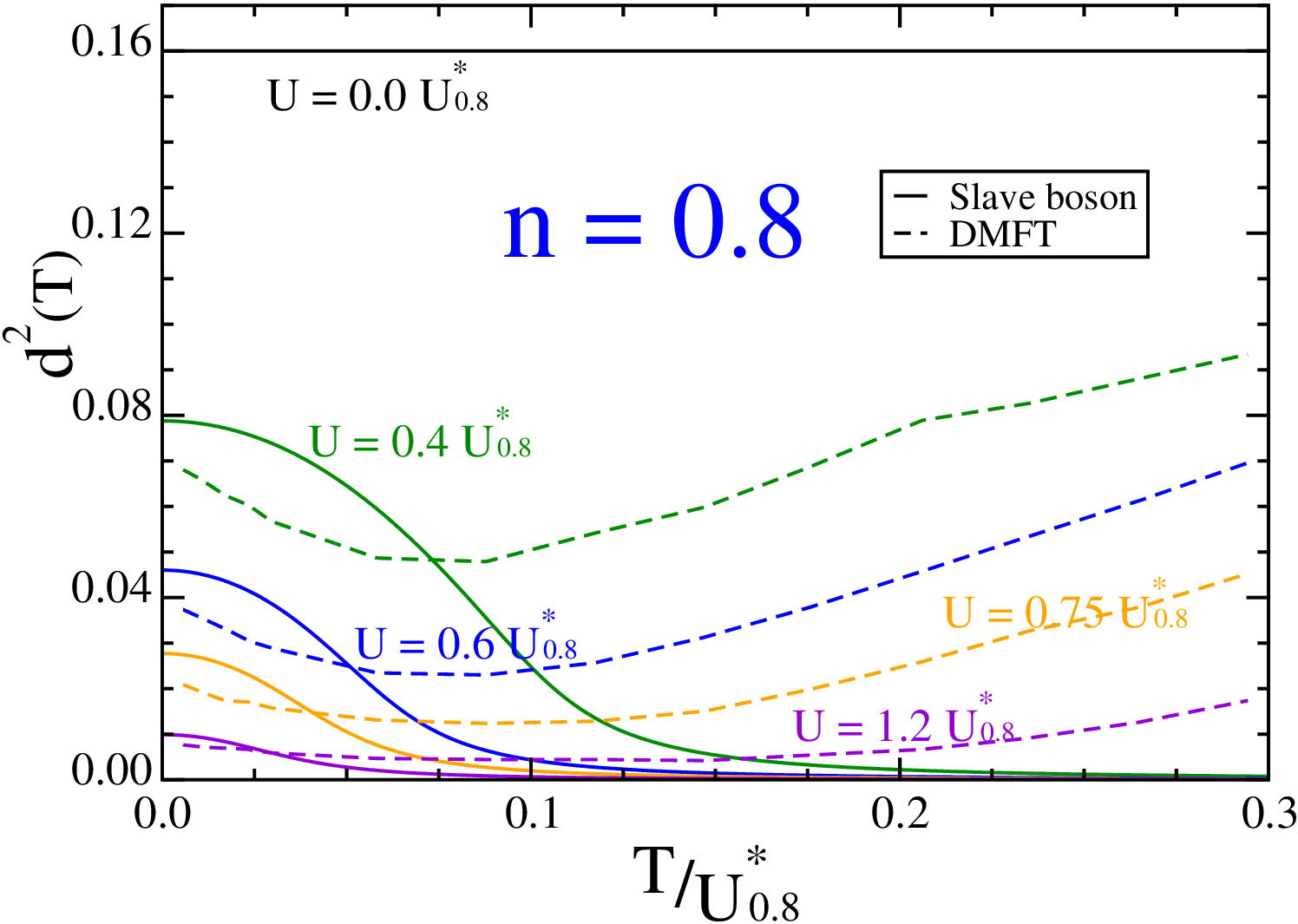}
\protect\caption{Same as Fig. \ref{n-100_d2} for $n=0.8$. Even though $d^2$ goes to zero at high $T$ the quasiparticle weight goes to a finite value because the system always remains metallic (cf. equation \eqref{equationq}).
\label{n-080_d2}}
\end{figure}
for different interaction strengths.
We use the results of Ref. [\onlinecite{Lanata2015a}]  based on DMFT simulations using the TRIQS\cite{Parcollet2015} library with a continuous time quantum Monte Carlo method\cite{Rubtsov2005,Werner2006} as the impurity solver for the Anderson impurity model associated with DMFT. Fig. \ref{n-080_d2} shows the same quantity for $n=0.8$. Good agreement is found in both Figs. for $T<T_\textrm{coh}$, and the DMFT results reproduce the physical behaviour of $d^2\rightarrow \frac{n^2}{4}$ when $T\rightarrow \infty$. We note the coincidence of our calculated $T_\textrm{coh}$ with the minima in the DMFT curves.
\begin{figure}[hbt]
\includegraphics*[width=0.46\textwidth,angle=0]{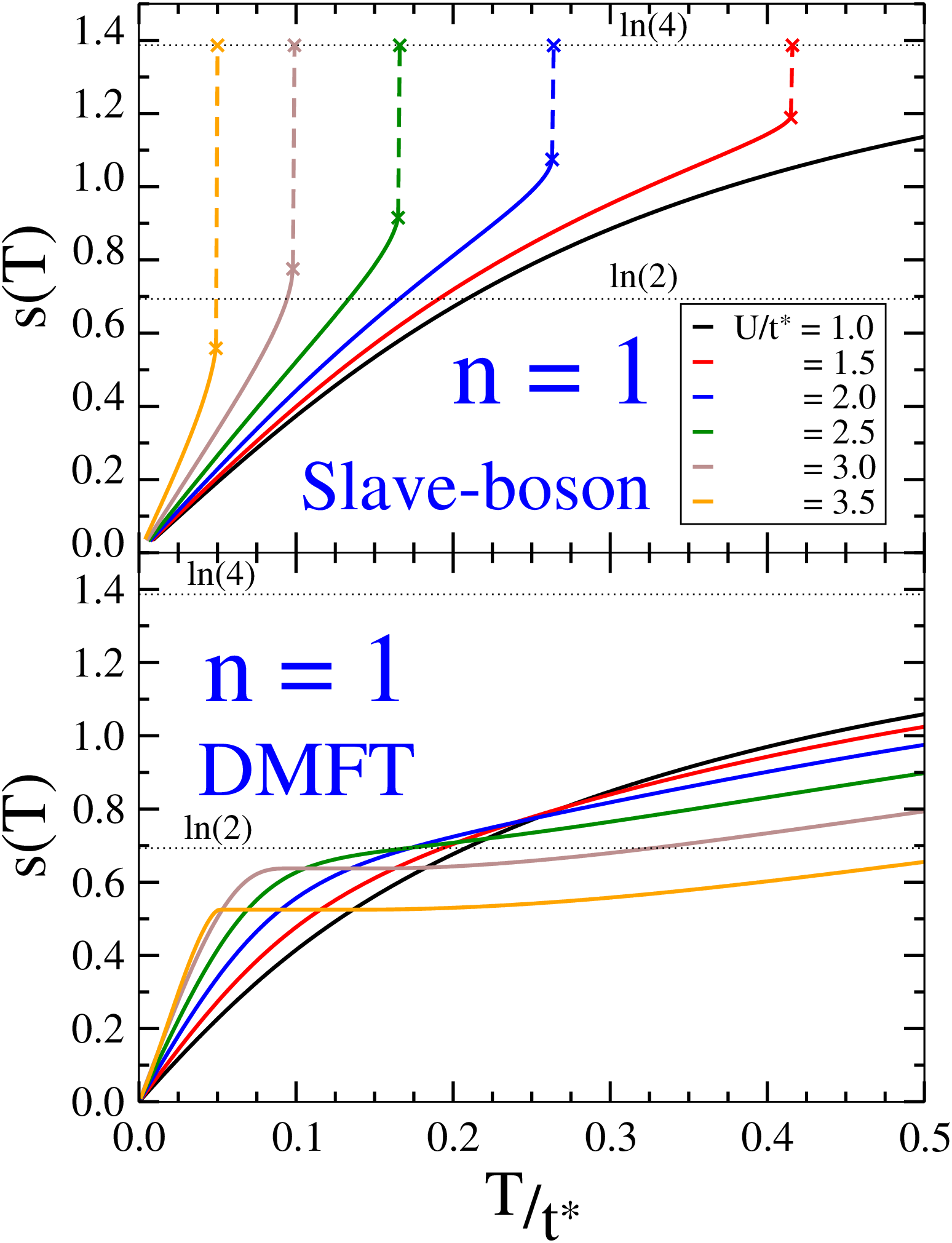}
\protect\caption{Entropy density at half-filling as function of temperature for different interaction values using the hypercubic lattice band structure for the slave-boson method (up) and DMFT from Ref. [\onlinecite{Pakhira2015b}] (bottom).
\label{n-100_S-hyper}}
\end{figure}

In Fig. \ref{n-100_S-hyper} we compare the entropy density at half-filling as a function of $T$ for the hypercubic lattice band structure calculated with our method and the DMFT results from Ref. [\onlinecite{Pakhira2015b}].
To make comparison easier, in this case we use $t^*$ as unit, where $U_c \simeq 4.5\, t^*$.
 In addition to the general behaviour described in Section \ref{entropy}, it is important to note that the reduction of the entropy value at the transition with increasing $U$ (end crosses at each solid line) is consistent with the appearance of the kink-like feature in the DMFT results.

\begin{figure}[hbt]
\includegraphics*[width=0.46\textwidth,angle=0]{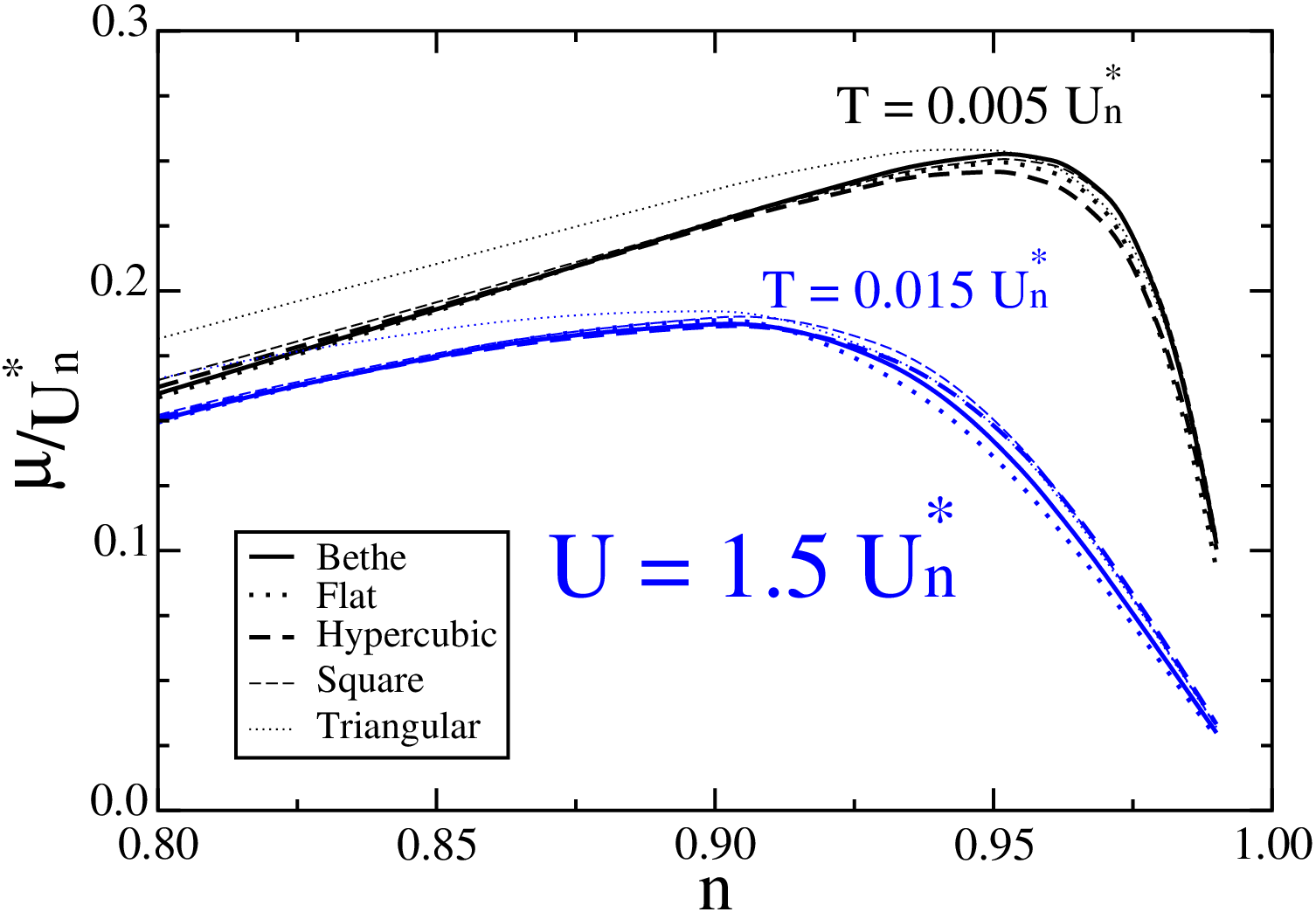}
\protect\caption{Dependence of the chemical potential $\mu$ with the filling number $n$ for $T/U^*_n= 0.005$ and $0.015$, calculated using different band structures. Note the weak dependence of $\mu$ over the various band structures. Regions with negative derivative define the shaded area in Fig. \ref{U-150_Tcoh}.
\label{U-150_mu}}
\end{figure}
Fig. \ref{U-150_mu} shows, for different band structures, the chemical potential as a function of the filling $n$ for temperatures $T/U^*_n = 0.005$ and $0.015$.
Segments with a negative slope of the curve determine, for each $T$, the region with negative charge compressibility.

%\bibliography{SlaveBosonPaper} % Tell bibtex which .bib file to use (this one is some example file in TexLive's file tree)

\begin{thebibliography}{72}
\bibitem {squarenote}%
  \BibitemOpen
   {
   It is well established that the ground state of the Hubbard model in the unfrustrated 
   square lattice, for any $U>0$, is a Mott insulator with antiferromagnetic correlations.
   This behaviour occurs due to a perfect nesting in the Fermi surface. In this paper
   we use paramagnetic solutions, which neglect such nesting effects, because we are mostly interested in the high$-T$ regime 
   where the Fermi liquid collapses and a crossover to a bad metal regime becomes possible.
   }
  \BibitemShut
  {NoStop}%
\end{thebibliography}

\end{document}